\documentclass[
    ,final            
  ]
  {aipproc}

\layoutstyle{6x9}

\begin{document}

\title{Phase-ordering kinetics: ageing and local scale-invariance}

\author{Malte Henkel}{
  address={Laboratoire de Physique des Mat\'eriaux (CNRS UMR 7556), 
  Universit\'e Henri Poincar\'e Nancy I, B.P. 239, 
  F - 54506 Vand{\oe}uvre-l\`es-Nancy Cedex, France}
}

\begin{abstract}
Dynamical scaling in ageing systems, notably in phase-ordering
kinetics, is well-established. New evidence in favour of Galilei-invariance
in phase-ordering kinetics is described.
\end{abstract}

\maketitle


\section{Dynamical scaling in ageing systems}

The study of the long-time dynamics of statistical systems far from equilibrium
has been a topic of intensive study. In many instances, the relaxation times
towards equilibrium can become extremely long such that the system stays for
all intents and purposes out of equilibrium. A paradigmatic example is 
provided by glassy systems which might be considered as an extremely viscous
liquid. In principle, the presence of very long relaxation time-scales might
suggest that quantitative properties of glassy dynamics should depend on 
a huge variety of microscopic `details' and furthermore, as the behaviour of
a glass may depend on its previous (thermal, mechanic,\ldots) history, those
system age. However, as first pointed out by
Struik in 1978 \cite{Stru78}, the ageing dynamics of many 
physically very different glass-forming
systems can be described in terms of {\em universal} master curves. The challenge is to try to understand the origin of this dynamical 
scale-invariance and to compute the form of these master curves from the 
essential characteristics of the system. 

Here we shall consider an analogous situation in supposedly more simple 
systems without intrinsic disorder or frustration. From some initial state
(typically one chooses a fully disordered initial state) the system is rapidly `quenched' either onto its critical point or else into a region of the phase diagram with at least two stable stationary states. The dynamics and an eventual
ageing behaviour (that is, a breaking of time-translation invariance) is then
observed. One distinguishes {\em physical ageing}, where the underlying
microscopic processes are reversible, from {\em chemico-biological ageing}, 
where irreversible microscopic processes may occur. In terms of models, 
a simple example for physical ageing is provided by the phase-ordering
kinetics of a simple Ising ferromagnet with purely relaxational dynamics 
quenched to below its critical temperature $T_c>0$. On the other hand, 
the ageing behaviour of the contact process (particles of a single species
$A$ move diffusively on a lattice and react according to $A+A\to\emptyset$ and
$A\to2A$) provides a paradigmatic case of ageing with underlying 
irreversible processes. 

Physically, these two kinds of ageing phenomena are quite distinct. In relaxing
ferromagnets, see e.g. \cite{Bray94,Cugl02,Henk04b} for reviews, 
there is for $T<T_c$ a non-vanishing surface tension between the
ordered domains which leads to the formation and growth of ordered clusters
of linear size $L=L(t)\sim t^{1/z}$, where $z$ is the dynamical exponent. 
For purely relaxational dynamics, it can be shown that $z=2$ for $T<T_c$ 
whereas for $T=T_c$, the non-trivial value of $z$ equals the one found for
equilibrium critical dynamics. If $\phi(t,\vec{r})$ denotes the time- and 
space-dependent order parameter, it is convenient to characterize the 
ageing behaviour in terms of the two-time correlation and linear response 
functions
\begin{equation}
C(t,s;\vec{r}) = \left\langle \phi(t,\vec{r})\phi(s,\vec{0})\right\rangle
\;\; , \;\;
R(t,s;\vec{r}) = \left. 
\frac{\delta\langle\phi(t,\vec{r})\rangle}{\delta h(s,\vec{0})}
\right|_{h=0}
\end{equation}
where $h$ is the magnetic field conjugate to $\phi$. The autocorrelation and
linear autoresponse functions are given by $C(t,s)=C(t,s;\vec{0})$ and
$R(t,s)=R(t,s;\vec{0})$. Here and later space-translation invariance will be
assumed. In phase-ordering, dynamical scaling is found in the regime where
\begin{equation} \label{agereg}
t\gg \tau_{\rm micro} \;\; , \;\;
s\gg \tau_{\rm micro} \;\; \mbox{\rm and} \;\;
t-s \gg \tau_{\rm micro}
\end{equation}
where $\tau_{\rm micro}$ is some microscopic reference time-scale. In the
ageing regime (\ref{agereg}) the only relevant lengths scales are describes
in terms of $L(t)$ and one expects
\begin{equation} \label{skCR}
C(t,s) = s^{-b} f_C(t/s) \;\; , \;\; 
R(t,s) = s^{-1-a} f_R(t/s) 
\end{equation}
with the asymptotics $f_{C,R}(y)\sim y^{-\lambda_{C,R}/z}$ for $y$ large. 
We stress the importance of the third condition in eq.~(\ref{agereg}) for the
validity of the scaling forms (\ref{skCR}). Throughout, the scaling limit
$t,s\to\infty$ with $y=t/s>1$ fixed will be implied. 
While the autocorrelation (autoresponse) exponents $\lambda_{C,R}$ are 
new, independent exponents, the exponents $a,b$ can be explicitly 
given.
At criticality $T=T_c$, one has $a=b=(d-2+\eta)/z=2\beta/\nu z$, 
where $\beta,\nu,\eta$ are
standard equilibrium critical exponents. For $T<T_c$, one has $b=0$ always. 
In simple scalar systems with short-ranged equilibrium correlations, such as 
the Ising or Potts models (with $d>1$), one has $a=1/z=1/2$. In the case of
long-ranged equilibrium correlations, $a$ may be different, i.e. $a=(d-2)/z$
in the $d$-dimensional spherical model. 

On the other hand, in irreversible ageing systems such as the critical
voter-model or the critical contact-process, there is no surface tension and
the dynamics proceeds through cluster dissolution \cite{Dorn01}. Still, in the 
ageing regime (\ref{agereg}) one observes again the same formal scaling 
behaviour eq.~(\ref{skCR}). However, the exponents $a$ and $b$ need no 
longer be the same even if one considers the ageing at a phase-transition in 
the non-equilibrium steady-state. For example, there is recent numerical evidence from the critical contact process in both $1D$ and $2D$ that
$b=2\beta/\nu_{\perp}z$ which naturally generalizes the result of critical
systems with detailed balance, but $a=b-1$ \cite{Enss04,Rama04}.

\section{Galilei-invariance in phase-ordering kinetics}

Having reviewed current knowledge on the dynamical scaling of ageing systems,
notably on the values of the ageing exponents $a$ and $b$, we now consider the
scaling functions $f_{C,R}(y)$ themselves. In particular, we ask whether
there exist any generic, model-independent argument which might inform us about
the form of the functions $f_{C,R}(y)$. In this context, the following
general statements about the dynamical symmetries of phase-ordering
kinetics can be made.
\begin{enumerate}
\item Time-translation invariance is broken.
\item There is dynamical scaling \cite{Bray94,Cugl02}, 
i.e. formally the order parameter satisfies the covariance condition
\begin{equation}
\phi(t,\vec{r}) = \alpha^{x_{\phi}} \phi(\alpha^{2} t,\alpha\vec{r})
\end{equation}
where $\alpha$ is a constant rescaling factor and $x_{\phi}$ an exponent. 
\item There is new evidence for Galilei-invariance which we now discuss. 
\end{enumerate}

Conventionally, one starts from a coarse-grained order parameter which 
is assumed to satisfy a Langevin equation, for example the one for model A
dynamics \cite{Bray94}
\begin{equation} \label{lang}
2{\cal M}\frac{\partial\phi}{\partial t} = \Delta \phi - 
\frac{{\rm d} V(\phi)}{{\rm d}\phi} + \eta
\end{equation}
where $\Gamma=(2{\cal M})^{-1}$ is a kinetic coefficient, $\Delta$ is the
spatial Laplacian, $V(\phi)$ is a typical double-well potential 
(e.g. $V(\phi)=(\phi^2-1)^2$) and $\eta$ is the thermal gaussian noise 
with covariance
$\langle\eta(t)\eta(t')\rangle=2 T\delta(t-t')$. In addition, one 
assumes a gaussian uncorrelated initial state with covariance
$\langle \phi(0,\vec{r})\phi(0,\vec{0})\rangle=a_0\delta(\vec{r})$. The
associated field-theoretic action in the Martin-Siggia-Rose (MSR) formalism is
\begin{equation} \label{msr}
S[\phi,\widetilde{\phi}]=S_0[\phi,\widetilde{\phi}]+S_b[\widetilde{\phi}]
\end{equation}
where $\widetilde{\phi}$ is the response field and 
\begin{eqnarray}
S_0[\phi,\widetilde{\phi}] &=& \int\!{\rm d}t{\rm d}\vec{r}\: 
\left[ \widetilde{\phi}(2{\cal M}\partial_t -\Delta)\phi +
\widetilde{\phi}\frac{\delta V(\phi)}{\delta \phi} \right] 
\nonumber \\
S_b[\widetilde{\phi}] &=& - T\int\!{\rm d}t{\rm d}\vec{r}\; 
\widetilde{\phi}^2(t,\vec{r}) - \frac{a_0}{2} \int\!{\rm d}\vec{r}\;
\widetilde{\phi}^2(0,\vec{r})
\end{eqnarray}
Then autocorrelators and autoresponses can be found as follows
\begin{equation}
C(t,s) = \left\langle \phi(t)\phi(s)\right\rangle \;\; , \;\;
R(t,s) = \left\langle \phi(t)\widetilde{\phi}(s)\right\rangle
\end{equation}

We shall consider as an extension of dynamical scaling the transformations
of the so-called Schr\"odinger group, defined by \cite{Nied72} 
\begin{equation}
t\to \frac{\alpha t+\beta}{\gamma t +\delta} \;\; , \;\;
\vec{r} \to \frac{{\cal R}\vec{r}+\vec{v}t+\vec{a}}{\gamma t+\delta}
\end{equation}
where $\alpha\delta-\beta\gamma=1$ and $\cal R$ is a $d$-dimensional rotation. 
Besides translations in time and space, rotations and scale transformations
with $z=2$, the Schr\"odinger group also contains the so-called
`special' transformations parametrized by $\gamma$. The Schr\"odinger group
is the maximal dynamical symmetry group of the free (and also of several non-linear) Schr\"odinger/diffusion equations and acts projectively
(i.e. up to a phase factor) on the wave function \cite{Nied72,Fush93}. 

The relationship of the Schr\"odinger group to phase-ordering kinetics 
can now be formulated in terms of the following three theorems. 

We consider an arbitrary space-time
infinitesimal coordinate transformation $\delta r_{\mu}=\epsilon_{\mu}$
with $\mu=0,1,\ldots,d$. Let $\eta$ stand for the phase picked up by the 
wave function $\phi$ under such a transformation. 
We call a MSR-theory {\em local} if the MSR-action (\ref{msr}) transforms as
\begin{equation}
\delta S= \int\!{\rm d}t{\rm d}\vec{r}\: 
\left( T_{\mu\nu}\partial_{\mu}\epsilon_{\nu} + J_{\mu}\partial_{\mu}\eta
\right) + \int_{(t=0)}\!{\rm d}\vec{r}\: 
\left( U_{\nu}\epsilon_{\nu} +V \eta\right)
\end{equation}
Here $T_{\mu\nu}$ is the energy-momentum tensor, $J_{\mu}$ a conserved
current and the second integral is only over the initial line $t=0$. 

\noindent {\bf Theorem 1:} {\it \cite{Henk03} For a local MSR-theory, one has}
\begin{equation}
\left. \begin{array}{l}
\mbox{\it phase-shift invariance} \\
\mbox{\it space-translation invariance} \\
\mbox{\it scale-invariance with $z=2$}\\
\mbox{\it Galilei-invariance}
\end{array} \right\} \Longrightarrow 
\mbox{\it special Schr\"odinger invariance}
\end{equation}
This result is completely analogous to a well-known result in conformal
field-theory. We point out that the requirement of time-translation invariance
is not required in order to derive the invariance under special 
Schr\"odinger transformations for local theories. 

To prove this, it suffices to write down the various 
infinitesimal transformations
explicitly. From invariance under phase shifts it follows $V=0$, 
space-translation invariance implies $U_1=\ldots=U_d=0$, from scale-invariance
it follows $2T_{00}+T_{11}+\ldots+T_{dd}=0$ and finally Galilei-invariance
implies $T_{0i}+2{\cal M}J_i=0$ for $i=1,\ldots,d$. Consequently
$\delta S=0$ under special Schr\"odinger transformations, as asserted. 
\hfill q.e.d. 

One may write down the tensor $T_{\mu\nu}$ and the current $J_{\mu}$
explicitly, for free fields this has been done in \cite{Henk03}. 

We now consider the r\^ole of Galilei-invariance more closely. It is 
well-known that a system coupled to a heat bath with a uniform temperature
$T>0$ cannot be Galilei-invariant. This can be easily seen from the
MSR-action (\ref{msr}), since the noise terms $S_b[\widetilde{\phi}]$ are not
invariant under a phase-shift $\phi\to e^{\eta}\phi$, $\widetilde{\phi}\to
e^{-\eta}\widetilde{\phi}$. At most, the deterministic part $S_0$ of the 
action might be Galilei-invariant. If that is the case, one has the 
following Bargman superselection rule
\begin{equation} \label{Barg}
\langle~ \underbrace{\phi\cdots\phi}_n  ~~
\underbrace{\widetilde{\phi}\cdots\widetilde{\phi}}_m ~\rangle_0 
\sim \delta_{n,m}
\end{equation}
The index $0$ refers to the average taken with respect to 
the deterministic part $S_0$ only.  

\noindent {\bf Theorem 2:} {\it \cite{Pico04} If the deterministic part $S_0$
of the MSR-action (\ref{msr}) is Galilei-invariant such that (\ref{Barg}) holds true, then} 
\begin{eqnarray}
R(t,s) &=& R_0^{(2)}(t,s) \\
C(t,s) &=& \frac{a_0}{2}\int\!{\rm d}\vec{r}\; 
R_0^{(3)}(t,s,0;\vec{r}) + T \int\!{\rm d}u{\rm d}\vec{r}\;
R_0^{(3)}(t,s,u;\vec{r})
\end{eqnarray}
{\it where the two-point function 
$R_0^{(2)}(t,s)=\langle\phi(t,\vec{y})\widetilde{\phi}(s,\vec{y})\rangle_0$ 
and the three-point function
$R_0^{(3)}(t,s,u;\vec{r})=\langle \phi(t,\vec{y})\phi(s,\vec{y})
\widetilde{\phi}^2(u,\vec{r}+\vec{y})\rangle_0$ are fixed by the
deterministic part $S_0$.} 

\noindent To see this, one merely has to include the noise term $S_b[\widetilde{\phi}]$ into the average, viz. $R=\langle\phi\widetilde{\phi}\rangle=
\langle\phi\widetilde{\phi}e^{-S_b[\widetilde{\phi}]}\rangle_0
=\langle\phi\widetilde{\phi}\rangle_0$ because of the Bargman superselection rule (\ref{Barg}). $C(t,s)$ is found similarly. \hfill q.e.d. 

Consequently, given the Galilei-invariance of the deterministic part, one
can find the contributions of both the thermal and the initial noise. 
Remarkably, the response function is noise-independent, whereas the correlator
vanishes in the absence of noise. 

Finally, we have to address the question whether the deterministic part of
the Langevin equation (\ref{lang}) can be Galilei-invariant. At first sight,
the answer seems to be negative, since a well-known mathematical fact
\cite{Fush93} states that the non-linear Schr\"odinger equation
\begin{equation}
\left(2m{\rm i}\partial_t - \Delta \right)\phi = F(t,\vec{r},\phi,\phi^*)
\end{equation}
is Schr\"odinger-invariant only for the special potential
$F=c(\phi\phi^*)^{2/d}\phi$, where $c$ is a constant. Furthermore, the 
solutions $\phi$ are necessarily complex. These difficulties can be
circumvented by considering the mass $m$ as a new dynamical variable
\cite{Giul96,Henk03}. We introduce a new wave function $\psi$ via
\begin{equation}
\phi(t,r) = \int_{-\infty}^{\infty}\!{\rm d}\zeta\; e^{-{\rm i}m\zeta}
\psi(\zeta,t,r)
\end{equation}
and look for the symmetries of the new non-linear `Schr\"odinger'-equation
\begin{equation} \label{schpsi}
\left(2\partial_{\zeta}\partial_t-\partial_r^2\right)\psi =
g F(\zeta,t,r,\psi,\psi^*)
\end{equation}
where we wrote the coupling constant $g$ of the non-linear term explicitly. 
Since from dimensional counting, $g$ is in general dimensionful, it should 
transform under the action of the Schr\"odinger group. We have systematically
constructed all such representations of the Schr\"odinger Lie algebra
and also of its subalgebra when time-translations are left out 
\cite{Stoi05}. We then find all invariant non-linear equations of the type
(\ref{schpsi}). The full analysis will be presented elsewhere, here 
we merely quote one special result. 

\noindent {\bf Theorem 3:} {\it \cite{Stoi05} If we write the solutions of
(\ref{schpsi}) $\psi=\psi_g(\zeta,t,r)=g^{(1-2x)/(4y)}\Psi(\zeta,t,r)$, where
$x$ is the scaling dimension of $\psi$ and $y$ the scaling dimension of $g$, then the equation}
\begin{equation} \label{stoi}
\left(2\partial_{\zeta}\partial_t-\partial_r^2\right)\Psi = g^{-5/4y}
f\left(g^{1/4y}\Psi\right) 
\end{equation}
{\it is Schr\"odinger-invariant, where $f$ is an arbitrary function.} 

We have obtained in this way a formulation of 
Schr\"odinger-invariance which allows for real-valued functions and 
furthermore should be flexible enough to include the double-well potentials
which enter into the Langevin equation (\ref{lang}). 

\section{Tests of Galilei-invariance}

The results quoted above in the theorems 1-3 
lead to quantitative predictions which have been
successfully tested in simulations. We first consider the response function,
which according to theorem~2 can be found from the assumption of its
covariant transformation under the Schr\"odinger group. This leads to
\cite{Henk03a,Pico04}
\begin{eqnarray}
R(t,s;\vec{r}) &=& R(t,s) 
\exp\left( -\frac{\cal M}{2}\frac{\vec{r}^2}{t-s}\right)
\;\; , \;\; 
R(t,s) = s^{-1-a} f_R(t/s) \nonumber \\
f_R(y) &=& f_0 y^{1+a'-\lambda_R/z} (y-1)^{-1-a'}
\label{RR}
\end{eqnarray}
where $\cal M$ and $f_0$ are non-universal constants. In most cases, one
has $a=a'$, but exceptions are known to occur, e.g. in the $1D$ Glauber-Ising
model. The gaussian space-dependence of $R(t,s;\vec{r})$ 
is characteristic of Galilei-invariance. Since response functions are
very much affected by noise and hence difficult to measure directly, a
convenient way of testing (\ref{RR}) is to study the spatio-temporally
integrated response
\begin{equation} \label{iR}
\int_0^s\!{\rm d}u \int_{0}^{\sqrt{\mu s}}\!{\rm d}r\: r^{d-1} R(t,u;\vec{r}) 
= r_0 s^{d/2-a} \rho^{(2)}(t/s,\mu) + \cdots 
\end{equation}
where $\mu$ is a control parameter, $r_0$ a constant related to $f_0$ 
and the function $\rho^{(2)}$ can be found explicitly from (\ref{RR})
\cite{Henk03a}. 

\begin{figure}
  \includegraphics[height=.3\textheight]{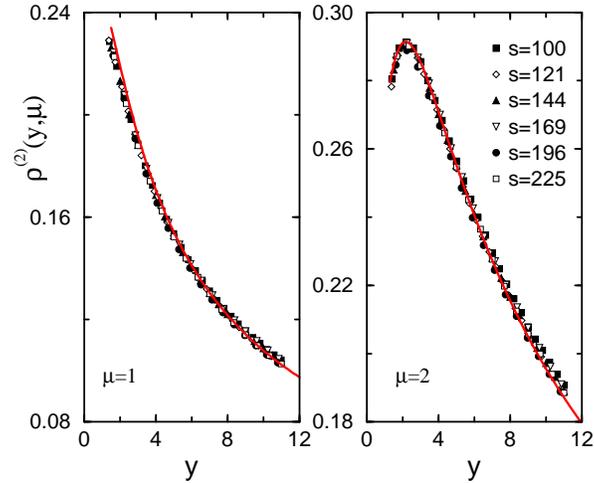}
  \caption{Scaling function $\rho^{(2)}(y,\mu)$ in the $2D$ Ising model
  at $T=0.66 T_c$ for two values of the control parameter $\mu$, as a function 
  of $y=t/s$ (after \cite{Henk03a}). \label{fig1}}
\end{figure}

In figure~\ref{fig1} we compare the prediction derived from (\ref{RR}) with
simulational data \cite{Henk03a} in the $2D$ kinetic Ising model, 
quenched into its ordered phase and with a purely relaxational heat-bath  dynamics. We used the exponents $z=2$, $a=a'=1/z=1/2$ 
and $\lambda_R=1.26$.
{}From the plots, one sees a nice collapse of the data obtained for 
several values of the waiting time $s$. Finally, the full curve agrees
perfectly with the data. Similar results also hold true in the $3D$ Ising
model \cite{Henk03a} 
and also for the $2D$ three-state Potts model \cite{Lore05}. 
Furthermore, the prediction (\ref{RR}) can be reproduced in the exactly
solvable $d$-dimensional spherical and $1D$ Glauber-Ising models. 

We remark that $R(t,s)$ is well reproduced in the critical $1D$ contact-process
\cite{Enss04}. 

As a second example, we consider the calculation of the autocorrelator
$C(t,s)$. From theorem 2, this requires the calculation of a 
noiseless three-point response function. 
Since Schr\"odinger-invariance fixes the three-point function only up
to an undetermined scaling function, a further argument is needed in order
to determine $C(t,s)$ completely. 
For models which are described by an underlying free-field theory, it
turns out that a simple heuristic idea based on the absence of singularities
in $C(t,s)$ leads to the following simple 
expression for $T=0$ in the scaling limit \cite{Pico04}
\begin{equation}
C(t,s) = f_C(t/s) \;\; , \;\; 
f_C(y) \approx C_0 \left( {(y+1)^2}/{y}\right)^{-\lambda_C/2}
\end{equation}
This agrees indeed with the exact solutions of the spherical model, the
spin-wave approximation of the XY model, the critical voter model and the free
random walk. 
It can also be checked \cite{Pico04} 
that the terms coming from thermal noise are 
irrelevant, as expected from renormalization group arguments \cite{Bray94}. 

Finally, for models not described by a free-field theory one can invoke
an extension of Schr\"odinger invariance to a new form of conformal 
invariance \cite{Henk04}. Then $C(t,s)$ can be written in terms of 
hypergeometric functions and this prediction again agrees nicely with
simulational data in both the $2D$ Ising \cite{Henk04} and three-states 
Potts models \cite{Lore05}.

Summarizing, we have presented evidence, both conceptual and simulational, 
which strongly suggest that
dynamical scaling in phase-ordering kinetics can be extended to the larger
Schr\"odinger group (without time-translations) of local scale-transformations. 
In particular, it appears that equations of the form (\ref{stoi}), 
rather than the Langevin equation (\ref{lang}) 
used traditionally, allow for a simple symmetry characterization. However, 
the derivation of equations of the type (\ref{stoi}) from a physical
argument remains to be understood. 

\noindent {\bf Acknowledgements:} It is a pleasure to thank A. Picone, M. Pleimling, S. Stoimenov and J. Unterberger for fruitful collaborations. 


%


\bibliographystyle{aipproc}   

\bibliography{granada8-henkel}

\IfFileExists{\jobname.bbl}{}
 {\typeout{}
  \typeout{******************************************}
  \typeout{** Please run "bibtex \jobname" to obtain}
  \typeout{** the bibliography and then re-run LaTeX}
  \typeout{** twice to fix the references!}
  \typeout{******************************************}
  \typeout{}
 }

\end{document}